\documentclass[conference]{IEEEtran}
\IEEEoverridecommandlockouts
\usepackage{cite}
\usepackage{bytefield}
\usepackage{amsmath,amssymb,amsfonts}
\usepackage{algorithmic}
\usepackage{graphicx}
\usepackage{wasysym}
\usepackage{textcomp}
\usepackage[table]{xcolor}
\usepackage{xcolor}
\usepackage{caption}
\usepackage{subcaption}
\usepackage{changepage}
\usepackage{url}
\usepackage{alltt}
\usepackage{graphicx}
\usepackage{balance}
\usepackage{booktabs}
\usepackage{soul}
\usepackage{float}

\def\BibTeX{{\rm B\kern-.05em{\sc i\kern-.025em b}\kern-.08em
    T\kern-.1667em\lower.7ex\hbox{E}\kern-.125emX}}

\usepackage[hidelinks]{hyperref}
\begin{document}

\title{CINNAMON: A Module for AUTOSAR Secure Onboard Communication\\
\thanks{}
}

\author{\IEEEauthorblockN{Giampaolo Bella, Pietro Biondi}
\IEEEauthorblockA{\textit{Dipartimento di Matematica e Informatica} \\
\textit{Universit\`a degli Studi di Catania}\\
Catania, Italy \\
giamp@dmi.unict.it,\\ pietro.biondi@phd.unict.it }
\and
\IEEEauthorblockN{Gianpiero Costantino, Ilaria Matteucci}
\IEEEauthorblockA{\textit{Istituto di Informatica e Telematica} \\
\textit{Consiglio Nazionale delle Ricerche}\\
Pisa, Italy \\
gianpiero.costantino@iit.cnr.it,\\ ilaria.matteucci@iit.cnr.it}
}

\maketitle

\begin{abstract}
This paper introduces CINNAMON, a software module that extends and seamlessly integrates with the AUTOSAR ``Secure Onboard Communication'' (SecOC) module~\cite{autosar:req,autosarR1911-2019} to also account for confidentiality of data in transit.
 
It stands for Confidential, INtegral aNd Authentic on board coMunicatiON (CINNAMON). It takes a resource-efficient and practical approach to ensure, at the same time, confidentiality, integrity and authenticity of frames. The main new requirement that CINNAMON puts forward is the use of encryption and thus, as a result, CINNAMON exceeds SecOC against \textit{information gathering} attacks. 

This paper sets forth the essential requirements and specification of the new module by detailing where and how to position it within AUTOSAR and by emphasizing the relevant upgrades with respect to SecOC. The presentation continues with the definition of a Security Profile and a summary of a prototype implementation of ours~\cite{DBLP:conf/codaspy/BellaBCM19,DBLP:conf/wisec/BellaBCM19}. While CINNAMON is easily extensible, for example through the definition of additional profiles, the current performances obtained on inexpensive boards support the claim that the approach is feasible. 

\end{abstract}


\section{Introduction}
Encryption is a \emph{de facto} standard to protect data while in transit. Many examples come in support of this claim, such as client-server communications over HTTPS and popular chat services pervasively exposed via mobile apps. Encryption is also one of the measures that (art. 32 of) the General Data Protection Regulation advocates to protect personal data in any application scenario that treats such data~\cite{gdpr}. The European Data Protection Board underlines ``\textit{the context of connected vehicles and mobility related applications}'' as one such scenario and explicitly calls for encryption \cite[\S2.7]{gdprauto}. It can be observed, however, that in-vehicle communications are not encrypted at present, and in fact the AUTOSAR SecOC module~\cite{autosar:req} only prescribes integrity and authenticity, as detailed below.

We contend that encryption ought to be added also to in-vehicle communications hence define CINNAMON, a module that requires encryption over the Controller Area Network (CAN). The pervasiveness of the applications of encryption outlined above somehow contributes to the motivation for our work. More importantly, a proof-of-concept of ours already demonstrated that the computational overhead can be negligible on currently inexpensive hardware~\cite{DBLP:conf/codaspy/BellaBCM19,DBLP:conf/wisec/BellaBCM19}, so feasibility increases our motivation significantly. Nonetheless, our main impulse is to counter \textit{information gathering} attacks, which may have dramatic consequences also at CAN bus level, such as the reverse engineering of proprietary mapping, normally stored in a \emph{Database CAN} (DBC) file~\cite{remoteattackJeep,DBLP:conf/cse/CostantinoM19}. A DBC stores the mapping between CAN frame payloads and functionalities of a vehicle, as decided by the Original Equipment Manufacturer (OEM). Therefore, a DBC is a clear aim for an attacker because it enables the Electronic Control Units (ECUs) of a specific vehicle to correctly interpret the payload values and translate them into signals that carry out the expected functionalities. Moreover, information gathering attacks may also be oriented at driver privacy infringement through profiling~\cite{bernardi_driver_2018,driverFingerprint}.

The number of ECUs in a modern vehicle ranges from a few tens to over a hundred.
To communicate with one another, ECUs may adopt several buses, such as the Controller Area Network~\cite{summacan}, FlexRay~\cite{flexray}, Ethernet~\cite{ethernet}.
In particular, the CAN bus is still the most widespread at present.
It leverages a binary language, standardised as ISO 11898-1:2015, to derive a simple protocol based on two bus lines~\cite{isocan}. However, CAN is not secure-by-design because authentication, integrity and confidentiality are not considered in the design and implementation of the protocol. This represents one of the main vulnerabilities of modern vehicles: getting (local or remote) access to CAN bus allows an attacker to inject unauthorised frames on the bus. 
These frames may compromise the functionalities of the target vehicle, eventually making them unsafe, as detailed below through the description of the threat model. There comes a non negligible inherent risk of, for example, malicious attacks against a vehicle, including remote control, as we shall detail in the next Section through real-world episodes.

AUTOSAR collects most of the strategies and guidelines that regulate the automotive world.
AUTOSAR depicts a Classic Platform, which is a Software Platform defined for deeply embedded systems and Application Software with high demands regarding predictability, safety and responsiveness.
~AUTOSAR covers functional safety and security aspects of onboard communications.
As for the latter, AUTOSAR proposes the \emph{Secure On Board Communication} Basic Software (BSW) module, named SecOC, listing its requirements~\cite{autosar:req} and providing its specification~\cite{autosarR1911-2019}. 
SecOC insists on the integrity of onboard communications and the authenticity of ECUs that act as senders. 
By contrast, it does not consider confidentiality.

This paper introduces the CINNAMON module, whose requirements leverage and extend those already provided by SecOC~\cite{autosar:req}. Therefore, 
CINNAMON is an AUTOSAR compliant Basic Software (BSW) module and insists not only on authenticity and integrity (as SecOC does) but also on confidentiality of CAN bus communications. 

CINNAMON aims at countering the information gathering (reverse engineering) activities of attackers by prescribing the symmetric-key encryption of frames that are in transit on the CAN bus.
In consequence, the attacker will not be able to read frame data fields unless she compromises at least one ECU to get the encryption key.
Thus, CINNAMON exceeds SecOC by design in the mitigation of information gathering attacks.

\smallskip
This paper is structured as follows. The next Section describes the assumed threat model. Section~\ref{sec:secoc} recalls the main features of SecOC. Section~\ref{sec:ourarchitecture} describes CINNAMON through its requirements, its specification and its security profiles.
Section~\ref{sec:motivation} provides an informal security assessment of the new module. Section~\ref{sec:RW} compares our contribution with the relevant literature, and Section~\ref{sec:conclusion} draws hints for future work and conclusions.

\section{Threat Model}\label{sec:threat}
Recent history about automotive security shows several examples of attacks to real vehicles.
In 2010, researchers showed how to control a car remotely, that is, make the car engine exploitable, turn off the brakes so that the vehicle would not stop, and make instruments give false readings~\cite{bbc}.
In 2015, hijacking was perpetrated on a Jeep Cherokee~\cite{remoteattackJeep} and also on a General Motors vehicle~\cite{GM2015}. Hackers  remotely took control of the engine and stole data from the infotainment system. They exploited the Internet connection of the infotainment system as well as a malicious version of the infotainment software installed on the car. 
In 2016, researchers hacked a TESLA Model S~\cite{Tesla2016} by using bugs on the TESLA's bounty program through which vehicles received firmware updates.
All these attacks leverage the lack of confidentiality for data in transit on the intra-vehicle CAN bus network, which are, consequently, exposed to several threats, hence the motivation for our work rises.
 
This paper assumes a threat model with an active attacker who may exploit some vulnerabilities of a car to gain some digital access to the car, either locally or remotely. 
More precisely, our attacker:
\begin{itemize}
\item may acquire pieces of information about the running protocols and other mechanisms in place in the network she observes,
\item may build and inject frames at will to manipulate the information processed by the target ECUs,
\item may not obtain privileged access to any ECUs.
\end{itemize}

Therefore, our threat model assumes the attacker to only have partial control of ECUs, hence she only has partial access to its functionalities. This would be the case, for example, when a Hardware Security Module (HSM) or similar solutions are adopted to protect cryptographic keys and run security-critical operations such as encryption.

In practice, the attacker may try to modify the behaviour of a target vehicle by sending customized CAN frames to trigger a specific functionality on a receiving ECU. The attacker in general \textit{aims to} mount the following attacks:
\begin{itemize}
	\item \emph{Replay}: re-use of valid CAN frames with malicious or fraudulent aims. 

	\item \emph{Tampering}: manipulation of CAN frames to spoil their contents so that a receiving ECU cannot perform the operation that was originally meant. 

	\item \emph{Forging}: generation of a valid CAN frame, which is then able to generate a valid signal and activate a specific ECU functionality.

	\item \emph{Fuzzing}: injection of CAN frames, which were previously forged, with the aim of studying the behaviour of a target ECU against unexpected inputs. 

	\item \emph{Masquerading}: misinterpretation of attacker's identity by using a CAN ID of some other genuine ECU, thereby masquerading as that ECU.
	\item \emph{Information Gathering}: identification of critical contents from CAN frames, such as the frame ID or payload and its associated ECU functionality, with the aim of using it against a target ECU to perform a post-attack.	
\end{itemize}

To mitigate such an attacker, we argue that a secure CAN protocol should achieve the following security properties:
\begin{itemize} 
	\item \textbf{Confidentiality.} The content of a frame is not disclosed to unauthorised entities.
	\item \textbf{Authentication.} The identity of the sender of a frame can be verified.
	\item \textbf{Integrity.} The content of a frame is not altered during transmission.
	\item \textbf{Freshness.} It can be verified whether a frame was already received.
\end{itemize}

\section{A primer on the Secure On Board Communication}\label{sec:secoc}
\begin{figure}[ht]
	\centering
	\includegraphics[scale=0.55]{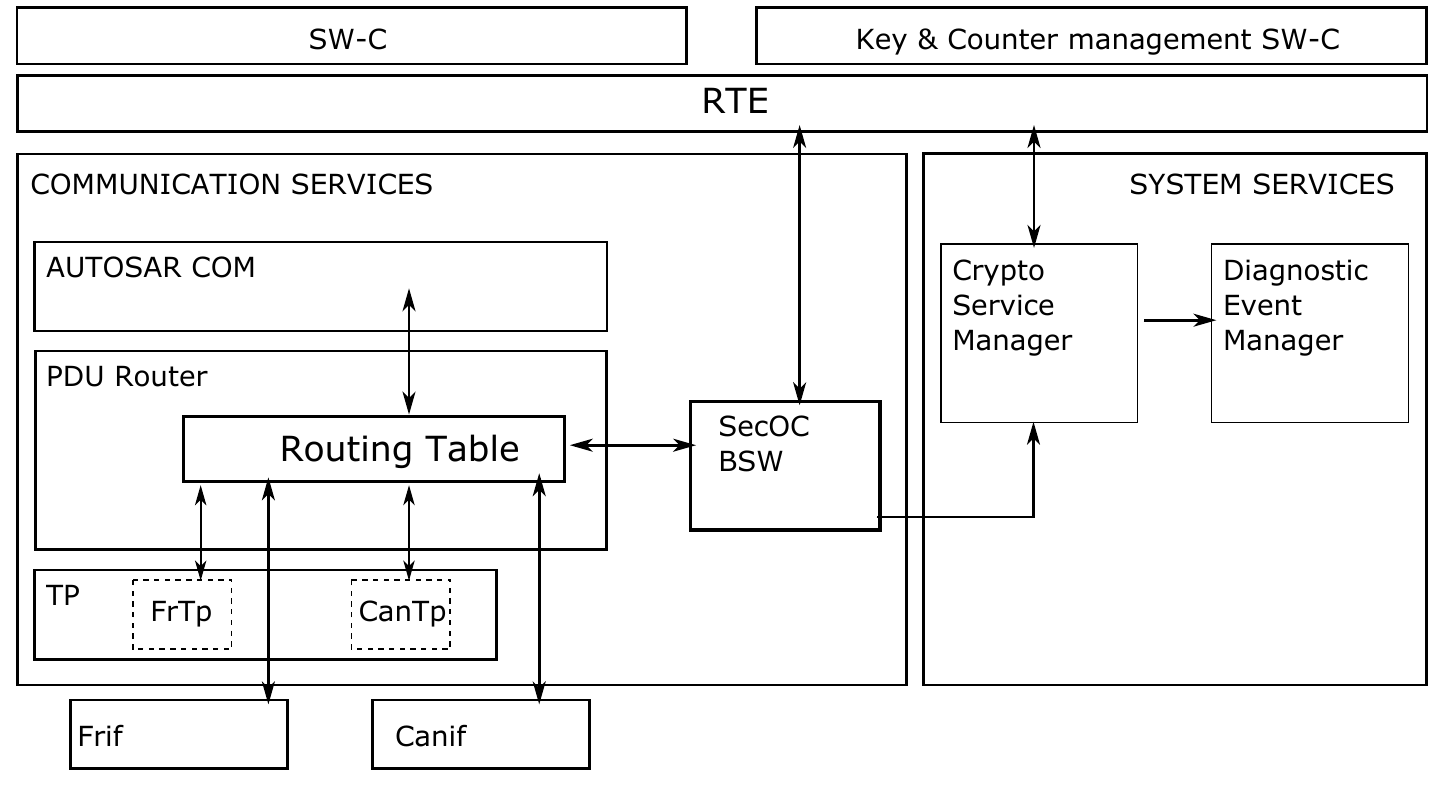}
	\caption{Integrating SecOC BSW module in AUTOSAR~\cite{autosarR1911-2019}}\label{fig:SecOCBSW}
\end{figure}

In the AUTOSAR Classic Platform, security of onboard communication is managed by a \emph{Basic Software Module}, SecOC, which is part of the \emph{Communication Services} (Fig.~\ref{fig:SecOCBSW}).
A Basic Software Module in AUTOSAR is a collection of documents to define a certain basic software functionality that may be deployed on an ECU.

AUTOSAR provides the list of all SecOC requirements~\cite{autosar:req}.
Such requirements regulate functional as well as non functional aspects.
More precisely, non functional requirements regulate the responsiveness of the module with respect to message communication.

\begin{figure}[ht]
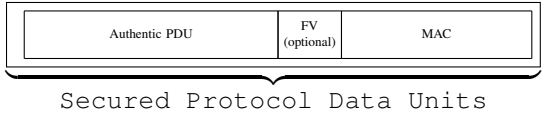

	\centering
	\begin{bytefield}[endianness=little,bitwidth=.30em]{64}
		$
		\underbrace{
			\framebox{
				\bitbox{32}{\tiny Authentic PDU}
				\bitbox{8}{\tiny FV (optional)}
				\bitbox{24}{\tiny MAC}}
		}_
		{\texttt{\normalsize Secured Protocol Data Units}}
		$
	\end{bytefield}
	\caption{Secured Protocol Data Units contents}\label{fig:SPDU}
\end{figure}

The SecOC module must be deployed over all communicating ECUs. In return, SecOC enables each ECU to verify the received messages in terms of authenticity and integrity through a Message Authentication Code (MAC) and in terms of freshness through the use of special counters, as we shall see below. Independently from the specific protocol, exchanged messages are in general addressed as Protocol Data Units (PDUs), thus
the SecOC Secure PDU can be depicted as in Fig.~\ref{fig:SPDU}.

As it is designed, the SecOC module is able to overcome all the attacks listed in \S\ref{sec:threat} except Information Gathering attacks, as the sequel of this paper demonstrates.

\section{The CINNAMON Module}\label{sec:ourarchitecture}
The CINNAMON BSW module leverages the SecOC design for what concerns authentication and integrity and adds confidentiality by introducing and regulating the use of cryptography. 
In this first version, CINNAMON intends to secure only the CAN bus, so PDUs can be concretely seen as CAN frames.

\subsection{Requirements}

This Section reviews the SecOC Requirements~\cite{autosar:req} and how they are inherited and upgraded in CINNAMON to fulfill the objectives of confidentiality, integrity and authenticity. Requirements that are inherited unaltered are normally not presented, except when they raise relevant discussion. Each requirement comes with a Status field representing to what extent the current prototype implementation covers it.

\smallskip
\paragraph{Functional}
CINNAMON\_00003 (Table~\ref{tab:req00003}) configures different security properties.
The security expert is in charge of defining the level of protection of onboard communication frames and the parameters needed to configure the functionalities of the module. 
\begin{table}[h]
	\centering
	\caption{CINNAMON\_00003}
	\scalebox{0.7}{
		\begin{tabular}{|l | l|}
			\hline
			\textbf{Functional Requirement} & \emph{Configuration of different security properites/requirements} \\ [0.5ex] 
			\hline\hline
			Type: & Valid\\
			\hline
			Description: & Different security properties, including confidentiality, shall be configurable.\\
			\hline
			Rationale: & The assessment may vary in several parameters and its security needs. \\&Thus the level of protection shall be configurable to adapt to these needs\\& by means of a set of adequate parameters.\\
			\hline
			Use Case: & Security experts define the different security properties, including\\& confidentiality. For every frame with security protection needs, \\&the appropriate properties may be selected. \\
			\hline
			Corresponding&\\
			SecOC Requirement & SRS\_SecOC\_00003\\
			\hline
			Status & Accomplished \\ [1ex]
			\hline
	\end{tabular}}	
	\label{tab:req00003}
\end{table}

CINNAMON satisfies this requirement by configuring confidentiality, authentication and integrity properties. Therefore,
in addition to the SecOC module, it also manages the parameters needed for encryption and decryption to work.
\smallskip
\paragraph{Initialisation}
CINNAMON\_00005 (Table~\ref{tab:req00005}) regulates also encryption and decryption functionalities.

With respect to SRS\_SecOC\_00005 requirement, the security configuration details in CINNAMON are not only restricted to Key-IDs for authentication and Freshness Values but also expand on the Key-IDs for encryption and decryption. 
\begin{table}[h]
	\centering
	\caption{CINNAMON\_00005}
	\scalebox{0.7}{
		\begin{tabular}{|l | l|}
			\hline
			\textbf{Initialisation} & \emph{Initialisation of security information} \\ [0.5ex] 
			\hline\hline
			Type: & Valid\\
			\hline
			Description: & CINNAMON security configuration shall get initialised at module start-up.\\
			\hline
			Rationale: & CINNAMON needs security configuration information e.g., Key-IDs for \\ & authentication, Freshness Values, and encryption/decryption to\\& perform its operations. Therefore, this information shall get recovered \\ &and configured before it starts its processing operation.\\
			\hline
			Use Case: & CINNAMON loads the ID of frames, the authorized authentication retry\\& counter, the used encryption mechanism and the properties that  are used \\&for the processing of its incoming communications from upper and lower layers.\\
			\hline
			Corresponding&\\
			SecOC Requirement & SRS\_SecOC\_00005\\
			\hline
			Status & Accomplished   \\ [1ex]
			\hline
	\end{tabular}}
	\label{tab:req00005}
\end{table} 

\smallskip
\paragraph{Normal Operations}
CINNAMON\_00012 (Table~\ref{tab:req00012}) requires the use of the module for different kinds of bus systems. It is a noteworthy requirement because, at the current stage, CINNAMON is only designed for the CAN bus and this is reflected by the `Partially accomplished' Status. We anticipate that further versions of our module could be extended for confidentiality on other protocol buses in the future.

\begin{table}[h]
	\centering
	\caption{CINNAMON\_00012}
	\scalebox{0.7}{
		\begin{tabular}{|l | l|}
			\hline
			\textbf{Normal Operations} & \emph{Support of Automotive BUS Systems}\\ [0.5ex] 
			\hline\hline
			Type: & Valid\\
			\hline
			Description: & CINNAMON shall be applicable for the different kind of bus systems that are\\& supported by AUTOSAR and that are typical for automotive environments.\\
			\hline
			Rationale: & All bus protocols supported by AUTOSAR shall benefit from the CINNAMON\\& design.\\
			\hline
			Use Case: & Low bandwidth buses like CAN shall be supported as well as technologies for\\& large data link, like Ethernet.\\
			\hline
			Corresponding&\\
			SecOC Requirement & SRS\_SecOC\_00012\\
			\hline
			Status & Partially accomplished  \\ [1ex]
			\hline
	\end{tabular}}
	\label{tab:req00012}
\end{table}

CINNAMON\_00030 (Table~\ref{tab:req00030}) requires the module to be able to extract the payload from secured frames, without authentication. In the current specification of CINNAMON this is true provided that frame decryption is performed first.
On the contrarily, SecOC is capable to directly extract the data payload.
\begin{table}[H]
	\centering
	\caption{CINNAMON\_00030}
	\scalebox{0.7}{
		\begin{tabular}{|l | l|}
			\hline
			\textbf{Normal Operations} & \emph{Support of capability to extract Authentic PDU without Authentication} \\ [0.5ex] 
			\hline\hline
			Type: & Valid\\
			\hline
			Description: & CINNAMON shall be capable to extract the payload from Secured \\&frames, without Authentication.\\
			\hline
			Rationale: & CINNAMON can be used as an extractor of payload from Secured frames,\\ & to enable low latency GW behaviour when a part of \\ & downstream communication clusters does not require authentication of frames.\\
			\hline
			Use Case: & Gateway.\\
			\hline
			Corresponding&\\
			SecOC Requirement & SRS\_SecOC\_00030 \\
			\hline
			Status & Partially accomplished   \\ [1ex]
			\hline
	\end{tabular}}
	\label{tab:req00030}
\end{table}

\smallskip
\paragraph{Support for end-to-end and point-to-point protection}
A point-to-point secure communication guarantees security between pairs of communicating ECUs. In case of a multi-hop communication, i.e., a frame circulates among several ECUs to reach a receiver ECU, each ECU authenticates the frame.
An end-to-end communication guarantees security only between the sender ECU and the receiver ECU regardless of the intermediate ECUs.
Frames exchanged in both communication types have to be protected.
\begin{table}[ht]
	\centering
		\caption{CINNAMON\_00013}
	\scalebox{0.61}{
		\begin{tabular}{|l | l|}
			\hline
			\textbf{Support for end-to-end and  point-to-point protection} & \emph{Support for end-to-end and point-to-point protection} \\ [0.5ex] 
			\hline\hline
			Type: & Valid\\
			\hline
			Description: & Support for end-to-end and point-to-point protection.\\
			\hline
			Rationale: & While some signals are simply forwarded and no further\\& requirements are given for the channel or relaying entities\\& in between, other may pass through relaying entities\\& that can do changes on the packet content and thus\\& need to be trusted by the receiving entity.\\
			\hline
			Use Case: & An ECU communicates data that is transmitted over several\\& logical networks with different security properties.\\& A re-authentication gateway bridges the data from a logical\\& network to the other and processes\\& verification and re-authentication.\\
			\hline
			Corresponding&\\
			SecOC Requirement & SRS\_SecOC\_00013\\
			\hline
			Status & Partially accomplished    \\ [1ex]
			\hline
	\end{tabular}}
	\label{tab:req00013}
\end{table}

CINNAMON aims at security on point-to-point communications.
This means that, if a CAN frame passes from multiple hops, all secure mechanisms have to be performed by each hop in the communication chain.
However, CINNAMON could be extended to obtain also end-to-end communication protection, by configuring the CINNAMON module so that an ECU does not act as receiver but forwards the frame.

\smallskip
\paragraph{Non-Functional Requirements}

CINNAMON\_00025 (Table~\ref{tab:req00025}) refers to computation performances.
Thus, it regulates the time needed to perform all security operations, notably considering also encryption and decryption.
\begin{table}[ht]
	\centering
		\caption{CINNAMON\_00025}
	\scalebox{0.7}{
		\begin{tabular}{|l | l|}
			\hline
			\textbf{Non-Functional Requirements (Timing)} & \emph{Authentication and verification processing time} \\ [0.5ex] 
			\hline\hline
			Type: & Valid\\
			\hline
			Description: & Authentication, verification, encryption and encryption\\&processing shall be performed in a timely fashion so \\&that the real time critical signals do not get affected.\\
			\hline
			Rationale: & Transmission and reception of the time critical frame \\& between the running applications of two or more peers \\& shall not get penalised by the additional processing of \\& their underlying communication software layers such that \\& the signals are finally rejected. It is necessary that when \\& time critical frames are transmitted and received through \\& a Secured frames, the additional processing required by \\&CINNAMON remains under a value that is predictable   \\& and compatible with the time constraints of the \\
			&concerned signals.\\
			\hline
			Use Case: & A legitimate authenticated frame is verified and passed to\\& the receiving ECU within the expected time-frame without \\&experiencing signal monitoring errors.\\
			\hline
			Corresponding&\\
			SecOC Requirement & SRS\_SecOC\_00025\\
		\hline
		Status & Accomplished  \\ [1ex]
			\hline
	\end{tabular}}
	\label{tab:req00025}
\end{table}

\subsection{Specification}
\begin{figure}[ht]
	\centering
	\includegraphics[scale=0.55]{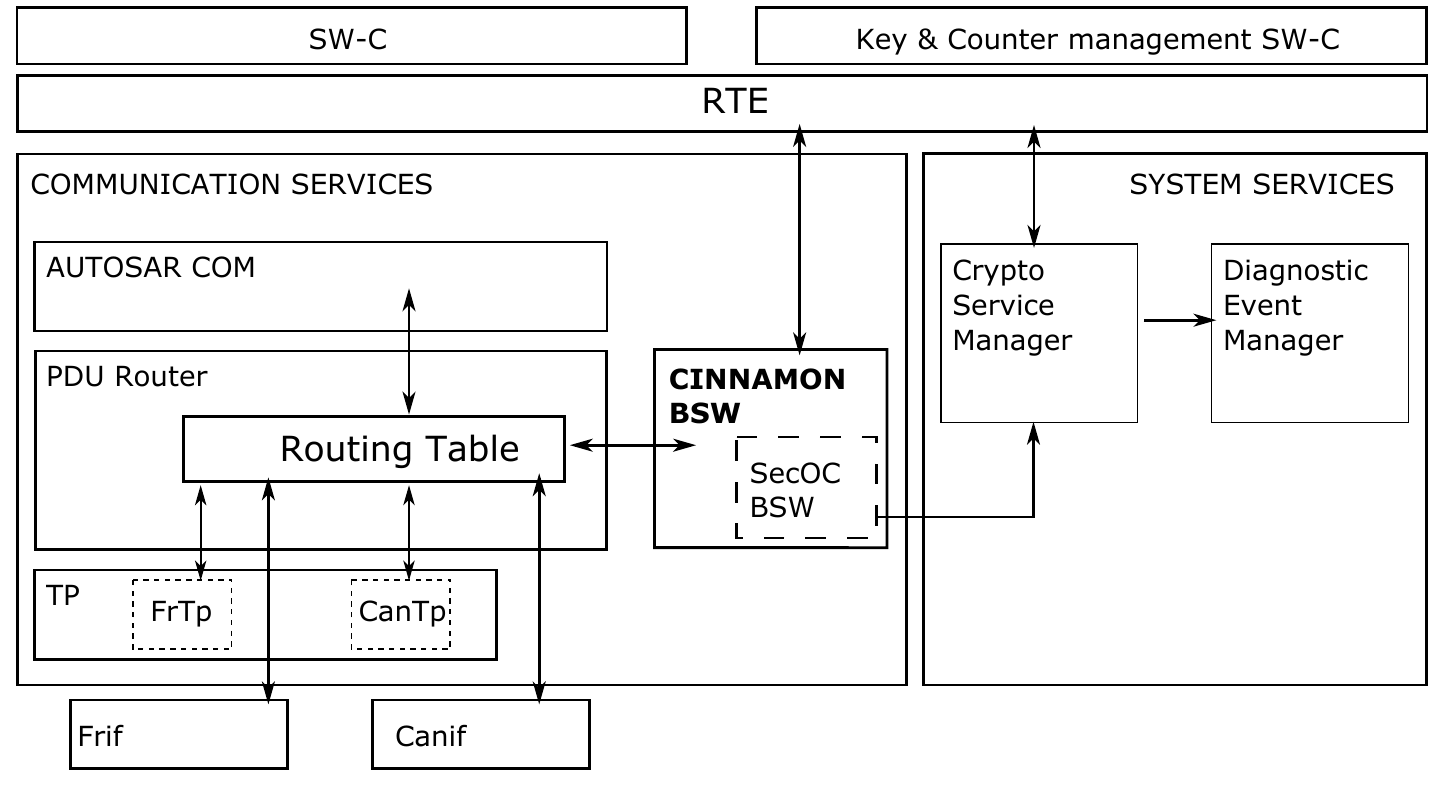}
	\caption{Integrating CINNAMON BSW module in AUTOSAR}\label{fig:ToucanModule}
\end{figure}

CINNAMON is, as SecOC, part of the Communication Services of the AUTOSAR Classic Platform, as depicted in Fig.~\ref{fig:ToucanModule}.
It encapsulates the SecOC module and inherits its API to interact with the PDU Router component (in this version to only manage CAN frames) and with the cryptographic services provided by the \emph{Crypto Service Manager}. Also, our module interacts with the Run-Time Environment to manage counters and keys.

CINNAMON acts as a middle-layer between the low-layer communication module, i.e., TP, and the upper layer software module, i.e., AUTOSAR COM. 
In addition, our module internally manages the communication with the lower level to build and send the secured data using a single CAN frame. Differently, the last version of SecOC specification~\cite{autosarR1911-2019} suggests to use two PDUs, one dedicated to store information used to authenticate the sender of the frame, and another one containing the secured frame.

\subsubsection{Authentication and Integrity}
\begin{figure}[ht]
	\centering
	\includegraphics[scale=0.5]{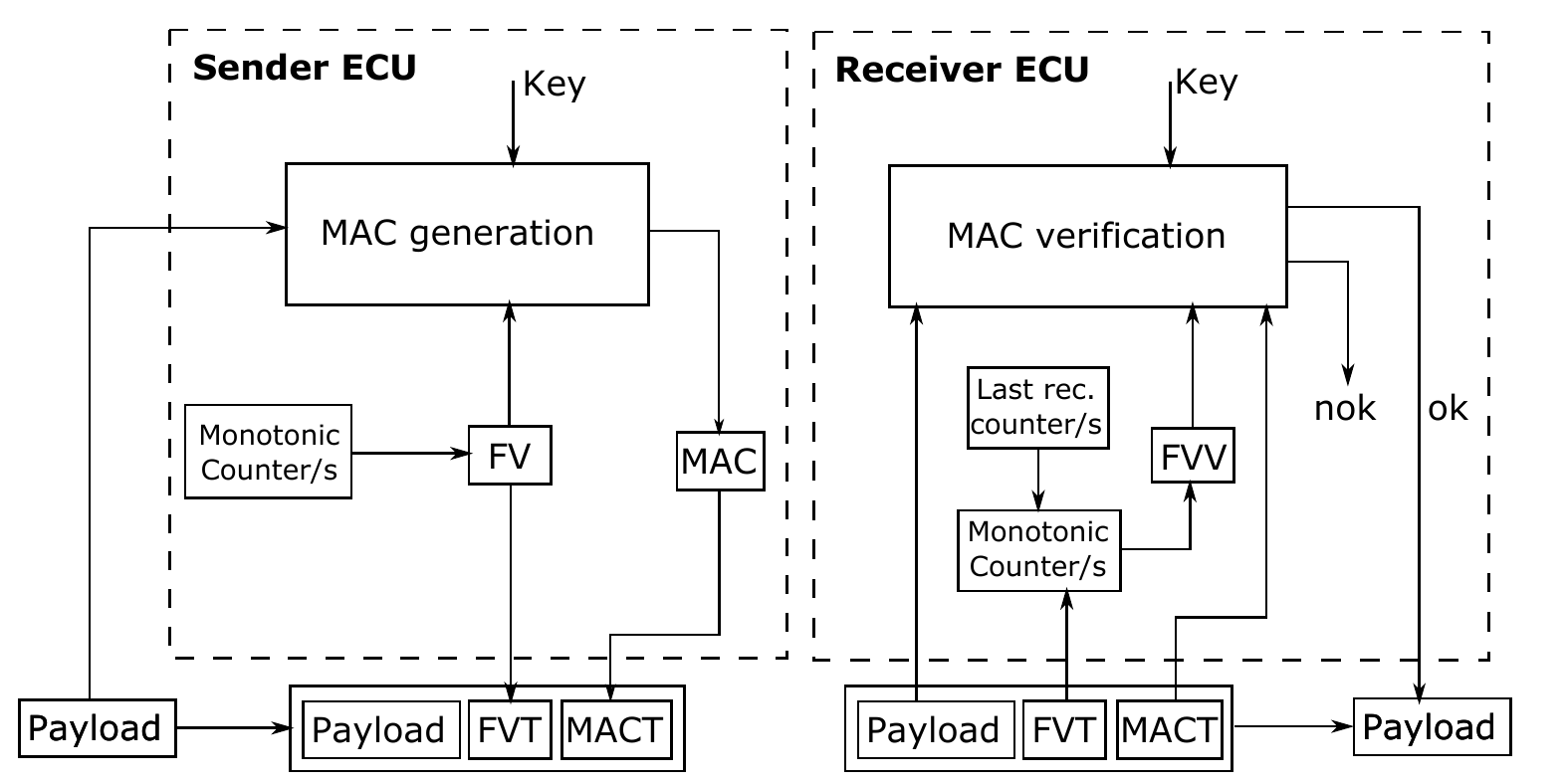}
	\caption{SecOc MAC Generation and Verification~\cite{autosarR1911-2019}}\label{fig:architectureAI}
\end{figure}

CINNAMON inherits SecOC authentication and integrity mechanisms, reviewed in Fig.~\ref{fig:architectureAI}.

AUTOSAR assumes that all ECUs have the cryptographic keys to handle Message Authentication Codes (MACs) (see~\cite{autosar:key}). Moreover, an external Freshness Manager provides counters to both sender and receiver to support the freshness of exchanged frames. 
\smallskip

CINNAMON inherits the same prerequisites, briefly recalled here. Let us consider a sender ECU and a receiver ECU. 
Before sending a payload, the sender  generates the MAC starting from the payload and possibly the \emph{Freshness Value} calculated according to the Monotonic Counter (Fig.~\ref{fig:architectureAI}) provided by the Freshness Manager (an ECU may decide to ignore the Freshness Value).
So, the secured CAN frame is composed by the payload, the truncated MAC (MACT in Fig.~\ref{fig:architectureAI}) and, optionally, the truncated freshness value (FVT).

The receiver has to validate the CAN frame before accepting it and does this by verifying the MAC.
In fact, the receiver generates a freshness value for verification (FVV) starting from the Monotonic Counter (Fig.~\ref{fig:architectureAI}) received by the Freshness Manager  and the previously received freshness value (the latest received counter in Fig.~\ref{fig:architectureAI}). Then, it calculates the MAC by using the received payload and the FVV. If the outcome equals 
the received MACT, then the payload is accepted, otherwise it is discarded.

\begin{figure*}[ht]
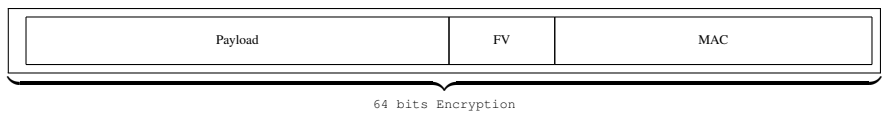

	\centering
	\begin{bytefield}[endianness=little,bitwidth=.50em]{64}
		$
		\underbrace{
			\framebox{
				\bitbox{32}{\tiny Payload}
				\bitbox{8}{\tiny FV}
				\bitbox{24}{\tiny MAC}}
		}_
		{\texttt{\tiny 64 bits Encryption}}
		$
	\end{bytefield}
	\caption{The CINNAMON Secured CAN Data field}\label{fig:securecanframe}
\end{figure*}

The CINNAMON module turns an AUTOSAR secured CAN frame into a CINNAMON secured CAN frame; its data field is presented in Fig.~\ref{fig:securecanframe}. 
A CINNAMON secured CAN frame is formed by reducing the dimension of the payload. Then, a freshness value is used to guarantee that the frame content is fresh. To complete the data field, an additional block is used for the Message Authentication Code (MAC), which ensures authentication and integrity. Finally, the entire 64 bits of the payload are encrypted to ensure confidentiality.

\subsubsection{Confidentiality}
CINNAMON aims at this property by taking the \emph{MAC-then-Encrypt} approach. 
Let $k$ represent the encryption key, $\tau$ the tag and $\nu$ the payload. What happens in operational terms is that the tag $\tau$ is appended to the payload $\nu$. Then, both the payload and the MAC are encrypted getting $C = ENC(k,\nu||\tau)$.  The opposite approach is \emph{Encrypt-then-MAC}, where $C = ENC(k,\nu)||\tau$, namely the MAC is calculated on the encrypted payload. 
Depending on the chosen algorithm and on the length of the frames, the \emph{MAC-then-Encrypt} approach may turn out less secure than the \emph{Encrypt-then-MAC} approach due to message padding, which may allow an attacker to break the security of the message rebuilding~\cite{Vaudenay}. However, this risk is zeroed in our case because there is no padding effect due to the fixed length of the considered messages.

There is a second reason in support of our choice. The \emph{MAC-then-Encrypt} approach encrypts 64-bit long frames (using encryption algorithms with 64-bit block size and no need for padding).
By contrast, using the \emph{Encrypt-then-MAC} approach according to the AUTOSAR specification, the payload (or payload plus FVT) is shorter than 64 bits hence padding would be needed. Most importantly, in frames where the 64 bits are already taken, adding a MAC would necessarily require the transmission of an additional frame to contain it~\cite{DBLP:conf/mtits/DarizSRCM17}.

A potential advantage of Encrypt-then-MAC could occur upon verification through the possibility to test the validity of the MAC as soon as the frame is received. However, this would not apply to CINNAMON because receiver needs to have the plain payload to validate the MAC.

\subsection{Security Profiles}
As in the Secure On Board Communication module, also in the CINNAMON module it is possible to define and manage various security profiles. 
\emph{Security profiles} provide a consistent set of values for a subset of parameters that are relevant for the configuration of CINNAMON~\cite{autosarR1911-2019}.
A CINNAMON security profile is defined as the configuration of the following mandatory parameters.
\begin{itemize}
	\item \small{\texttt{algorithmFamily:String [0..1]}} \normalsize is the first parameter that characterises the used authentication algorithm. This parameter identifies the family of authentication algorithms.
	\item \small{\texttt{algorithmMode:String [0..1]}} \normalsize is the second parameter that characterises the used authentication algorithm. This parameter identifies which MAC algorithm of the family is used.
	\item \small{\texttt{algorithmSecondaryFamily:String [0..1]}} \normalsize is the third parameter that characterises the used authentication algorithm. This parameter identifies a secondary family of authentication algorithms, if any.
	\item \small{\texttt{authInfoTxLength:PositiveInteger}} \normalsize denotes the length of the truncated MAC.
	\item \small{\texttt{freshnessValueLength:PositiveInteger}} \normalsize denotes the length of the generated freshness value.
	\item \small{\texttt{freshnessValueTruncLength:PositiveInteger}} \normalsize denotes the length of the truncated freshness value inserted in a frame.
	\item \small{\texttt{algorithmFreshnessValue:String [0..1]}} \normalsize denotes the algorithm used to generate the freshness value.
	\item \small{\texttt{algorithmEncryption:String [0..1]}} \normalsize denotes the encryption algorithm.
\end{itemize}

Note that the first six parameters are inherited from SecOC, while the last two are typical of CINNAMON.

This paper defines one example security profile, given in Table~\ref{tab:profile1}. It can be seen that it requires 24 bits for the truncated MAC, and this is coherent with the choice of SecOC. It relies on Chaskey MAC, which is robust under tag truncation~\cite{chaskey14} and on SPECK64/128, a lightweight block cipher publicly released by the NSA~\cite{speck13}.

It is clear that the very definition of this security profile was influenced by feedback from our experiments, as we shall see (\S\ref{sec:toucan}).

\begin{table}[h]
	\centering
	\caption{Example CINNAMON Security Profile}
	\scalebox{1.1}{
		\begin{tabular}{l c}
			\hline
			Parameter & Configuration Value \\ [0.5ex] 
			\hline\hline
			\texttt{algorithmFamily} & Chaskey\\
			\texttt{algorithmMode} & Chaskey\_MAC\\
			\texttt{algorithmSecondaryFamily} & not set\\
			\texttt{SecOCFreshnessValueLength} & not set\\
			\texttt{SecOCFreshnessValueTruncLength} & not set\\
			\texttt{SecOCAuthInfoTruncLength} & 24 bit\\
			\texttt{algorithmFreshnessValue} & not set\\
			\texttt{algorithmEncryption} & SPECK64/128 \\ [1ex]
			\hline
	\end{tabular}}
	\label{tab:profile1}
\end{table}

\section{Security Assessment}\label{sec:motivation}
This section presents an informal security assessment of CINNAMON in comparison to SecOc. We assume a scenario in which an attacker does not have total control of ECUs, hence she does not have privileged access to the boards and cannot see the cryptographic keys each board stores, for example protected under an HSM. The assessment can be compactly represented as in Table~\ref{tab:compareporperty}, which relates the security properties, and in Table~\ref{tab:comparethreat}, which displays mitigation of threats.
\begin{table}[ht]
	\centering
	\caption{Security Properties}
	\scalebox{1}{	
		\begin{tabular}{|c | c| c|}
			\hline
			\textbf{Security Property} & \textbf{SecOC} & \textbf{CINNAMON}\\ [0.5ex] 
			\hline\hline
			Confidentiality & \Circle  & \CIRCLE\\
			\hline
		Authentication   & \CIRCLE & \CIRCLE\\
			\hline
			Integrity  &\CIRCLE  & \CIRCLE \\
			\hline
			Freshness  &\CIRCLE  & \CIRCLE \\ [1ex]
			\hline
	\end{tabular}}
	\label{tab:compareporperty}
\end{table}

\begin{table}[ht]
	\centering
	\caption{Mitigated Threats}
		\scalebox{1}{
	\begin{tabular}{|c | c| c|}
		\hline
		\textbf{Threats} & \textbf{SecOC} & \textbf{CINNAMON}\\ [0.5ex] 
		\hline\hline
		Replay & \CIRCLE  & \CIRCLE\\
		\hline
		Tampering  & \CIRCLE & \CIRCLE\\
		\hline
		Forging  &\CIRCLE  & \CIRCLE \\ 
		\hline
		Fuzzing  &\CIRCLE  & \CIRCLE \\
		\hline
		Masquerading &\CIRCLE  & \CIRCLE \\
		\hline
		Information Gathering &\Circle  & \CIRCLE\\
		[1ex]
		\hline
	\end{tabular}}	
	\label{tab:comparethreat}
\end{table} 

It is visible that CINNAMON counters information gathering by aiming at the confidentiality property. However, this must be further spelled out. 
It means that a module implementation that is SecOC compliant still brings an appreciable inherent risk that an attacker sniffs CAN frames, attempts to interpret them and continues with fuzzing to observe the reactions of target ECUs.
The attacker may ultimately learn the semantics of each observed frame and, eventually, re-create the whole DBC of the vehicle. Remarkably, this could be possible even without any access to the MAC keys because frames are sent in the clear. CINNAMON minimises such risk.

On the other hand, we concede that, should the attacker manage to fully compromise an ECU gaining access to all its stored keys, both modules would become ineffective. However, this scenario is thwarted at another level, that of hardware-based security, an area of ferment at least in the last two decades. As a start, the already mentioned HSMs, such as ARM TrustZone~\cite{ARMtrustzone}, protect the memory areas that stores keys. We could even conjecture several levels of increasing hardware protection (and costs) depending on the specific protection requirements of the given application scenario. For example, the encryption key could be stored in a TPM 2.0 and the MAC key in an HSM, an architecture that would cause multiple violation efforts to the attacker.

\section{Outline of prototype implementation}\label{sec:toucan}

A prototype implementation of the CINNAMON security profile seen above is available. Its main features and performances are outlined here, while the full details are published elsewhere~\cite{DBLP:conf/codaspy/BellaBCM19,DBLP:conf/wisec/BellaBCM19}.

Coherently with the definition of the profile, no parameters related to freshness are used, so the secured frame consists of 40 bits for the payload and 24 bits for the MAC. Encryption is the outermost operation, as it was required.

\subsection{Testbed}
Our prototype is deployed on two STM32F407 Discovery boards, each with an ARM Cortex M4 processor. The boards provide physical input buttons, plus light emitting diodes for visual outputs. Both are connected to a workstation via a USB-to-CAN interface.

\subsection{Implementation complexities}
Our initial choice to use Chaskey as a MAC function upon the basis of its specifications was a lucky one. The function was reasonably easy to implement and appreciably fast since the initial experiments. Tags were truncated to 24 bits.

However, the encryption algorithm had to be chosen with care. Our obvious, initial candidate was AES but it produced a data field of at least 128 bits, while we aimed at a data field of 64 bits only, coherently with Fig.~\ref{fig:securecanframe}, so that it could be accommodated in just one frame. On the other hand, a 64 bit version of AES would be weaker and is not standardised. We also experimented with DES, 3DES and Blowfish, but their main drawback for our application was the computational overhead. By contrast,
SPECK64/128 uses a 128 bit key, produces a 64 bit output and is lightweight, so it turns out the optimal candidate here.

\subsection{Demonstration}

The sender board is in charge of sending secured frames. The board uses a 128 bit key to calculate the MAC on the given 40 bit payload and then truncates the resulting tag to 24 bits to complete the allowed 64-bit data field. After that, the payload is encrypted with SPECK 64/128 and finally sent to the other board over the bus.

When the other board receives the secured frame, it first decrypts the entire payload of 64 bits, then it calculates the MAC on the 40 bit payload. It truncates the live MAC to 24 bits and compares it with the 24 bits of the MAC received from the sending board. If all checks succeed, then the board turns on the blue led, otherwise it turns on the red led.

\subsection{Performances}
The additional computational overhead due to the management of MACs and of encryption could be a deal breaker. As said above, it was inevitable to proceed by trial and error, with some candidate encryption schemes that had to be abandoned. However, we were pleased to observe that handling the Chaskey/SPECK pair only negligibly reflects on the overall performance. Despite the use of inexpensive hardware with clock at 168 MHz, we face an average of under 6$\mu$s to generate or interpret a secured frame.
\rm

\section{Related Work}\label{sec:RW}
The current literature has several contributions around the lack of security of the CAN bus. Some solutions refer to the software level, others to the protocol level.

Among the software solutions is TACAN~\cite{TACAN}, a covert-channel based solution. It shares a master key between an ECU and the Monitor Node to generate shared session keys. 
These are assumed to be stored in a tamper-resistant memory of a security module, such as the Trusted Platform Module (TPM)~\cite{isoTPM}. 
Each ECU embeds unique authentication frames into CAN frames and continuously transmits them through covert channels, which can be received and verified by the Monitor Node. 
TACAN aims at mitigating suspension, injection and masquerade attack. 
With respect to TACAN, CINNAMON also mitigates  other attacks, such as information gathering and fuzzing attack. 

Authentication has been most considered. It is the aim of CANAuth~\cite{canauth11}, based on CAN+~\cite{canplus09}, which is an extension of the basic CAN protocol, and of MaCAN~\cite{MaCAN12}, a centralized authentication protocol based on broadcast-based authentication. 
LCAP~\cite{LCAP12} is a protocol for frame authentication, resistance to replay attacks, and backward compatibility at the same time. 
Libra-CAN~\cite{libra12} is a protocol based on a MAC calculated using MD5.
CaCAN~\cite{CaCAN14} also introduces a key distribution phase inherited from existing protocols.
The protocol needs a new component to be inserted in the vehicular network in order to act as a monitoring node. Frames are not sent in broadcast but on a peer-to-peer base.
In LeiA~\cite{RaduG16}, for each frame, the protocol sends a frame in plain-text and another one with the MAC of the frame.

CANcrypt \cite{cancrypt} is closely related to our work but is not AUTOSAR compliant. Also TLS-based approaches are valid but demand extra-vehicular Internet connectivity and are limited to time-critical applications due to performance overhead \cite{TLSoverCAN}.

The general limitation of all these approaches is the lack of integration with the AUTOSAR platform and, in particular, with the the current SecOC module. By contrast, we have seen that the CINNAMON module integrates with AUTOSAR and enables protocols to be implemented in compliance with its security profiles.

A framework for the specification and automatic generation of security features for communications among AUTOSAR-compliant components must be mentioned~\cite{autosarComponentBerna}. It allows AUTOSAR designers to add security specifications to the communication model through a dedicated software tool. 
However, it has not yet been practically used to advance new components or protocols that would combine confidentiality with authentication and integrity.

\section{Conclusions and Future Work}\label{sec:conclusion}
CINNAMON is an AUTOSAR compliant basic software module for confidentiality, integrity and authenticity on CAN bus.
It is designed as an enhancement of the Secure Onboard Communication (SecOC) module~\cite{autosarR1911-2019} and inherits its freshness mechanisms. The main distinctive feature is that CINNAMON  manages confidential CAN communication by encrypting CAN frames, hence it effectively thwarts information gathering attacks.

CINNAMON is scalable in the sense that it can host additional security profiles that become necessary depending on the application domain. It is reassuring that the current profile has reached the level of a prototype implementation whose performances are promising on inexpensive hardware. 
As ongoing work, we are defining new security profiles and are working on a possible implementation for each.
As future work, we plan to extend CINNAMON to secure not only the CAN bus but also other buses, such as CAN-FD and Ethernet.
This picture supports the claim that CINNAMON is feasible.

\section*{ACKNOWLEDGEMENT}
This work has been partially supported by the COSCA research project (NGI\_TRUST 2nd Open Call (ref: NGI\_TRUST 2019002)).

\bibliographystyle{plain}
\balance
\bibliography{automotive}
\end{document}